\pdfoutput=1
\documentclass[12pt,a4paper]{article}

\usepackage{ifthen} 
\newboolean{pdflatex}
\setboolean{pdflatex}{true} 

\newboolean{articletitles}
\setboolean{articletitles}{true} 

\newboolean{uprightparticles}
\setboolean{uprightparticles}{false} 

\def\paperauthors{LHCb collaboration} 
\def\paperasciititle{Ratio of Lbar to KS production cross-sections in pp collisions at 13.6 TeV}
\def\papertitle{Ratio of \Lbar to \KS production cross-sections in $pp$ collisions at 13.6\tev} 
\def\paperkeywords{{High Energy Physics}, {LHCb}} 
\def\papercopyright{\the\year\ CERN for the benefit of the LHCb collaboration} 
\def\paperlicence{CC BY 4.0 licence}
\def\paperlicenceurl{https://creativecommons.org/licenses/by/4.0/}

\newif\ifEnableSectionTOCLinks
\EnableSectionTOCLinksfalse 

\usepackage[top=1in, bottom=1.25in, left=1in, right=1in]{geometry}

\usepackage{microtype}
\usepackage{lineno}  
\usepackage{xspace} 
\usepackage{caption} 

\usepackage{graphicx}  
\usepackage{color}
\usepackage{colortbl}
\graphicspath{{./figs/}} 

\usepackage{amsmath} 
\usepackage{amssymb}
\usepackage{amsfonts}
\usepackage{upgreek} 

\usepackage[normalem]{ulem} 

\newcommand*\patchAmsMathEnvironmentForLineno[1]{%
\expandafter\let\csname old#1\expandafter\endcsname\csname #1\endcsname
\expandafter\let\csname oldend#1\expandafter\endcsname\csname
end#1\endcsname
 \renewenvironment{#1}%
   {\linenomath\csname old#1\endcsname}%
   {\csname oldend#1\endcsname\endlinenomath}%
}
\newcommand*\patchBothAmsMathEnvironmentsForLineno[1]{%
  \patchAmsMathEnvironmentForLineno{#1}%
  \patchAmsMathEnvironmentForLineno{#1*}%
}
\AtBeginDocument{%
\patchBothAmsMathEnvironmentsForLineno{equation}%
\patchBothAmsMathEnvironmentsForLineno{align}%
\patchBothAmsMathEnvironmentsForLineno{flalign}%
\patchBothAmsMathEnvironmentsForLineno{alignat}%
\patchBothAmsMathEnvironmentsForLineno{gather}%
\patchBothAmsMathEnvironmentsForLineno{multline}%
\patchBothAmsMathEnvironmentsForLineno{eqnarray}%
}

\usepackage[pdftex,
            pdfauthor={\paperauthors},
            pdftitle={\paperasciititle},
            pdfkeywords={\paperkeywords}]{hyperref}
\usepackage{hyperxmp}
\hypersetup{
    pdfcopyright={Copyright (C) \papercopyright},
    pdflicenseurl={\paperlicenceurl}
}

\usepackage[colorinlistoftodos,textsize=scriptsize]{todonotes}

\usepackage[bottom,flushmargin,hang,multiple]{footmisc}

\usepackage[all]{hypcap} 

\usepackage{xspace} 
\usepackage{upgreek}

\def\lhcb   {\mbox{LHCb}\xspace}

\def\MagUp {\mbox{\em Mag\kern -0.05em Up}\xspace}

\ifdefined\ifuprightparticles
\else
\newboolean{uprightparticles}
\setboolean{uprightparticles}{false} 
\fi

\ifthenelse{\boolean{uprightparticles}}%
{

 \def\Ppi         {\ensuremath{\uppi}\xspace}

 \def\PDelta      {\ensuremath{\Delta}\xspace}                 
 \def\PXi         {\ensuremath{\Xi}\xspace}                 
 \def\PLambda     {\ensuremath{\Lambda}\xspace}                 
 \def\PSigma      {\ensuremath{\Sigma}\xspace}                 
 \def\POmega      {\ensuremath{\Omega}\xspace}                 
 \def\PUpsilon    {\ensuremath{\Upsilon}\xspace}
 \let\oldPi\Pi
 \def\PPi         {\ensuremath{\oldPi}\xspace}

 \def\PB      {\ensuremath{\mathrm{B}}\xspace}                 
 \def\PD      {\ensuremath{\mathrm{D}}\xspace}                 

 \def\PK      {\ensuremath{\mathrm{K}}\xspace}                 
 \def\Pb      {\ensuremath{\mathrm{b}}\xspace}                 
 \def\Pc      {\ensuremath{\mathrm{c}}\xspace}

 \def\Pp      {\ensuremath{\mathrm{p}}\xspace}                 
 \def\Pq      {\ensuremath{\mathrm{q}}\xspace}                 
 \def\Ps      {\ensuremath{\mathrm{s}}\xspace}

 \def\thebaroffset{0.0em}
}
{

 \def\Ppi         {\ensuremath{\pi}\xspace}

 \mathchardef\PDelta="7101
 \mathchardef\PXi="7104
 \mathchardef\PLambda="7103
 \mathchardef\PSigma="7106
 \mathchardef\POmega="710A
 \mathchardef\PUpsilon="7107
 \mathchardef\PPi="7105
 \def\PB      {\ensuremath{B}\xspace}                 
 \def\PD      {\ensuremath{D}\xspace}                 

 \def\PK      {\ensuremath{K}\xspace}                 
 \def\Pb      {\ensuremath{b}\xspace}                 
 \def\Pc      {\ensuremath{c}\xspace}

 \def\Pp      {\ensuremath{p}\xspace}                 
 \def\Pq      {\ensuremath{q}\xspace}                 
 \def\Ps      {\ensuremath{s}\xspace}

 \def\thebaroffset{0.18em}
}
\newcommand{\offsetoverline}[2][\thebaroffset]{\kern #1\overline{\kern -#1 #2}}%

\makeatletter
\ifcase \@ptsize \relax
  \newcommand{\miniscule}{\@setfontsize\miniscule{4}{5}}
\or
  \newcommand{\miniscule}{\@setfontsize\miniscule{5}{6}}
\or
  \newcommand{\miniscule}{\@setfontsize\miniscule{5}{6}}
\fi
\makeatother

\DeclareRobustCommand{\optbar}[1]{\shortstack{{\miniscule (\rule[.5ex]{1.25em}{.18mm})}
  \\ [-.7ex] $#1$}}

\def\quark     {{\ensuremath{\Pq}}\xspace}
\def\quarkbar  {{\ensuremath{\overline \quark}}\xspace}

\def\squark    {{\ensuremath{\Ps}}\xspace}

\def\cquark    {{\ensuremath{\Pc}}\xspace}

\def\bquark    {{\ensuremath{\Pb}}\xspace}

\def\pion   {{\ensuremath{\Ppi}}\xspace}

\def\pip    {{\ensuremath{\pion^+}}\xspace}
\def\pim    {{\ensuremath{\pion^-}}\xspace}

\def\kaon    {{\ensuremath{\PK}}\xspace}

\def\KorKbar {\kern \thebaroffset\optbar{\kern -\thebaroffset \PK}{}\xspace}

\def\KS      {{\ensuremath{\kaon^0_{\mathrm{S}}}}\xspace}
\def\Vzero   {{\ensuremath{V^0}}\xspace}

\def\D       {{\ensuremath{\PD}}\xspace}

\def\DorDbar {\kern \thebaroffset\optbar{\kern -\thebaroffset \PD}\xspace}

\def\Dp      {{\ensuremath{\D^+}}\xspace}
\def\Dm      {{\ensuremath{\D^-}}\xspace}

\def\DpDm    {\ensuremath{\Dp {\kern -0.16em \Dm}}\xspace}

\def\B       {{\ensuremath{\PB}}\xspace}

\def\BorBbar {\kern \thebaroffset\optbar{\kern -\thebaroffset \PB}\xspace}

\def\Bd      {{\ensuremath{\B^0}}\xspace}

\def\BdorBdbar {\kern \thebaroffset\optbar{\kern -\thebaroffset \Bd}\xspace}

\def\Bs      {{\ensuremath{\B^0_\squark}}\xspace}

\def\BsorBsbar {\kern \thebaroffset\optbar{\kern -\thebaroffset \Bs}\xspace}

\def\Y#1S{\ensuremath{\PUpsilon{(#1S)}}\xspace}

\def\proton      {{\ensuremath{\Pp}}\xspace}
\def\antiproton  {{\ensuremath{\overline \proton}}\xspace}

\def\Lz          {{\ensuremath{\PLambda}}\xspace}
\def\Lbar        {{\ensuremath{\offsetoverline{\PLambda}}}\xspace}
\def\LorLbar     {\kern \thebaroffset\optbar{\kern -\thebaroffset \PLambda}\xspace}

\newcommand{\decay}[2]{\ensuremath{\mathinner{#1\!\to #2}}\xspace}

\def\to                 {\ensuremath{\rightarrow}\xspace}

\def\AT#1     {\ensuremath{A_{\mathrm{T}}^{#1}}\xspace}           

\def\C#1      {\ensuremath{\mathcal{C}_{#1}}\xspace}                       
\def\Cp#1     {\ensuremath{\mathcal{C}_{#1}^{'}}\xspace}                    
\def\Ceff#1   {\ensuremath{\mathcal{C}_{#1}^{\mathrm{(eff)}}}\xspace}        
\def\Cpeff#1  {\ensuremath{\mathcal{C}_{#1}^{'\mathrm{(eff)}}}\xspace}       
\def\Ope#1    {\ensuremath{\mathcal{O}_{#1}}\xspace}                       
\def\Opep#1   {\ensuremath{\mathcal{O}_{#1}^{'}}\xspace}                    

\newcommand{\aunit}[1]{\ensuremath{\text{\,#1}}}       

\newcommand{\tev}{\aunit{Te\kern -0.1em V}\xspace}
\newcommand{\gev}{\aunit{Ge\kern -0.1em V}\xspace}
\newcommand{\mev}{\aunit{Me\kern -0.1em V}\xspace}
\newcommand{\kev}{\aunit{ke\kern -0.1em V}\xspace}
\newcommand{\ev}{\aunit{e\kern -0.1em V}\xspace}
 
\newcommand{\mevc}{\ensuremath{\aunit{Me\kern -0.1em V\!/}c}\xspace}
\newcommand{\gevc}{\ensuremath{\aunit{Ge\kern -0.1em V\!/}c}\xspace}
\newcommand{\mevcc}{\ensuremath{\aunit{Me\kern -0.1em V\!/}c^2}\xspace}
\newcommand{\gevcc}{\ensuremath{\aunit{Ge\kern -0.1em V\!/}c^2}\xspace}

\def\mm   {\aunit{mm}\xspace}

\def\barn{\aunit{b}\xspace}

\newcommand{\chisq}{\ensuremath{\chi^2}\xspace}

\def\gsim{{~\raise.15em\hbox{$>$}\kern-.85em
          \lower.35em\hbox{$\sim$}~}\xspace}
\def\lsim{{~\raise.15em\hbox{$<$}\kern-.85em
          \lower.35em\hbox{$\sim$}~}\xspace}

\def\pt         {\ensuremath{p_{\mathrm{T}}}\xspace}

\def\evtgen     {\mbox{\textsc{EvtGen}}\xspace}

\def\geant      {\mbox{\textsc{Geant4}}\xspace}

\def\photos     {\mbox{\textsc{Photos}}\xspace}

\def\pythia     {\mbox{\textsc{Pythia}}\xspace}

\def\tell1  {TELL1\xspace}
\def\ukl1   {UKL1\xspace}

\newcommand{\ie}{\mbox{\itshape i.e.}\xspace}

\newcommand{\lhcborcid}[1]{\href{https://orcid.org/#1}{\hspace*{0.1em}\raisebox{-0.45ex}{\includegraphics[width=1em]{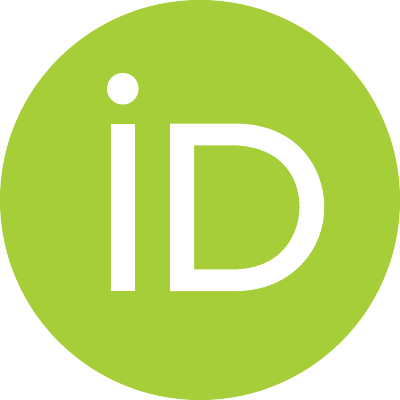}}}}

\hypersetup{
  colorlinks   = true, 
  urlcolor     = blue, 
  linkcolor    = blue, 
  citecolor    = red   
}

\ifEnableSectionTOCLinks
    \usepackage[explicit]{titlesec} 
    
    \let\oldcontentsline\contentsline
    \renewcommand

    \titleformat{\section}{\normalfont\Large\bf}{\hyperlink{tocsection.\thesection}{{\thesection} \parbox[t]{\dimexpr\textwidth-1pc}{#1}}}{1pc}{}

    \titleformat{\subsection}{\normalfont\bf}{\hyperlink{tocsubsection.\thesubsection}{{\thesubsection} \parbox[t]{\dimexpr\textwidth-1pc}{#1}}}{1pc}{}

    \titleformat{name=\section,numberless}[display]{}{}{0pt}{\normalfont\Huge\bfseries #1}
\fi

\usepackage{cite} 
\usepackage{LHCb/mciteplus}
\makeatletter
\g@addto@macro\bfseries{\boldmath}
\makeatother

\usepackage{longtable} 
\usepackage{rotating}
\usepackage{amsmath}
\usepackage{float}
\usepackage{booktabs}
\usepackage{siunitx}
\usepackage{subcaption}

\begin{document}

\renewcommand{\thefootnote}{\fnsymbol{footnote}}
\setcounter{footnote}{1}


\begin{titlepage}
\pagenumbering{roman}

\vspace*{-1.5cm}
\centerline{\large EUROPEAN ORGANIZATION FOR NUCLEAR RESEARCH (CERN)}
\vspace*{1.5cm}
\noindent
\begin{tabular*}{\linewidth}{lc@{\extracolsep{\fill}}r@{\extracolsep{0pt}}}
\ifthenelse{\boolean{pdflatex}}
{\vspace*{-1.5cm}\mbox{\!\!\!\includegraphics[width=.14\textwidth]{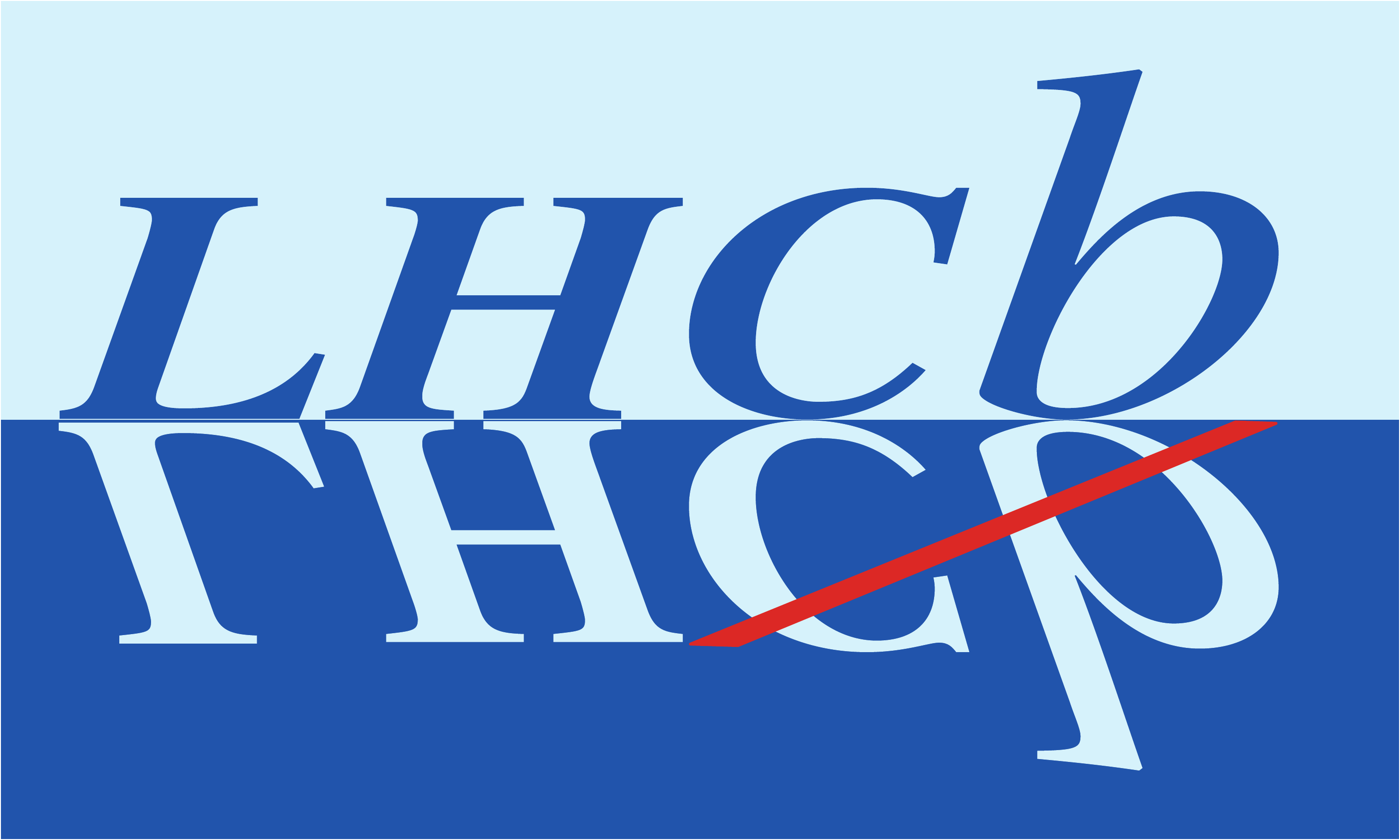}} & &}%
{\vspace*{-1.2cm}\mbox{\!\!\!\includegraphics[width=.12\textwidth]{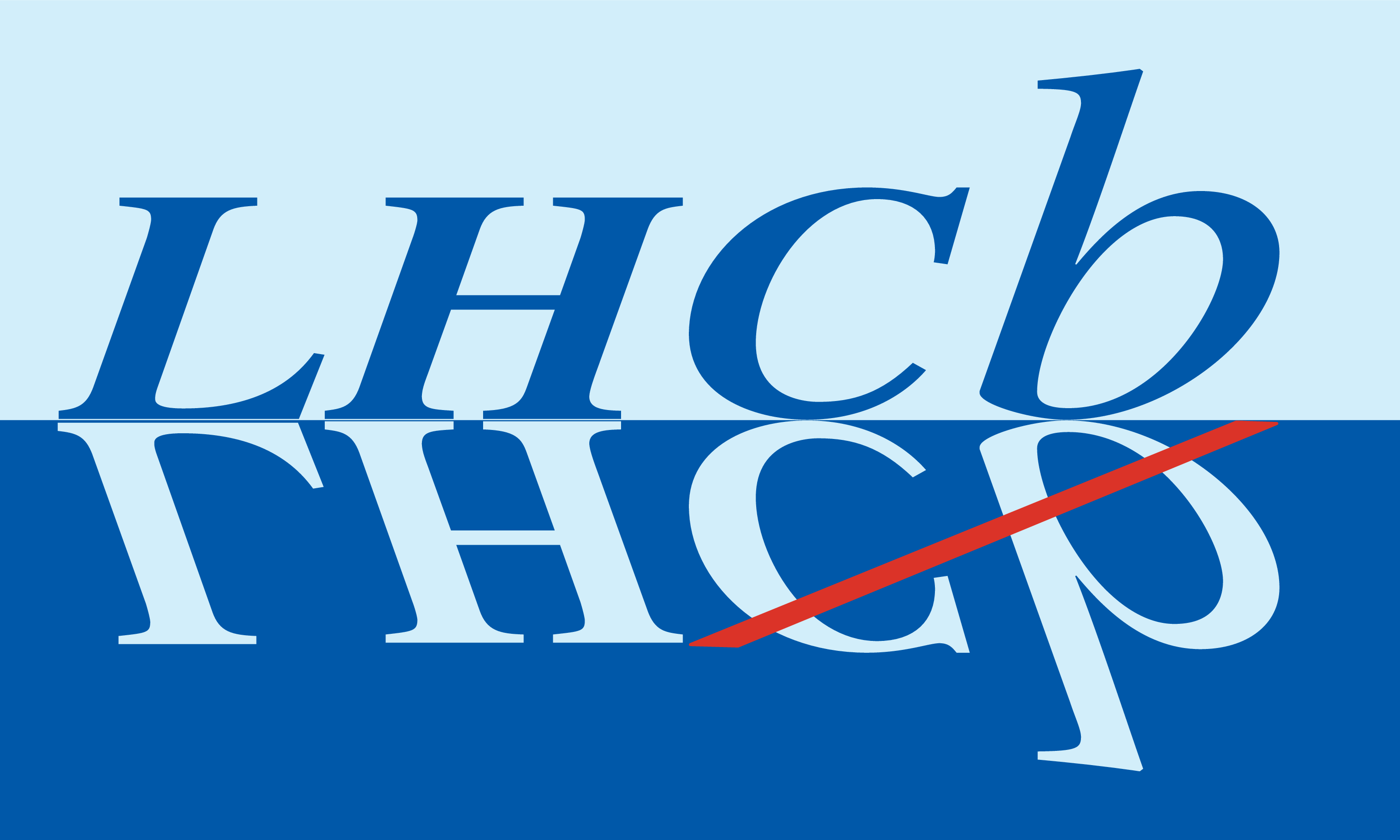}} & &}%
\\
 & & CERN-EP-2026-297 \\  
 & & LHCb-PAPER-2026-045 \\  
 & & \today \\ 
 & & \\
\end{tabular*}

\vspace*{4.0cm}

{\normalfont\bfseries\boldmath\huge
\begin{center}
  \papertitle 
\end{center}
}

\vspace*{2.0cm}

\begin{center}
\paperauthors\footnote{Authors are listed at the end of this paper.}
\end{center}

\vspace{\fill}

\begin{abstract}
  \noindent
  The production cross-section ratio of promptly produced \Lbar and \KS hadrons is measured in proton-proton collision data at a centre-of-mass energy of 13.6\tev recorded in 2024 with the upgraded LHCb detector and corresponding to an integrated luminosity of $\SI{64}{\pico\barn^{-1}}$. The ratio is presented as  a function of transverse momentum, \pt, and rapidity, $y$, in the ranges $0.15 < \pt < 2.5$\gevc and $2.0 < y < 4.5$. The results are in good agreement with previous measurements at different centre-of-mass energies. They are compared to predictions from currently available event generators and can be used as input in future event generator tunings.   
  
\end{abstract}

\vspace*{2.0cm}

\begin{center}
  Submitted to
  JHEP 
\end{center}

\vspace{\fill}

{\footnotesize 
\centerline{\copyright~\papercopyright. \href{\paperlicenceurl}{\paperlicence}.}}
\vspace*{2mm}

\end{titlepage}


\newpage
\setcounter{page}{2}
\mbox{~}
%
%
%
%


\renewcommand{\thefootnote}{\arabic{footnote}}
\setcounter{footnote}{0}


\cleardoublepage


\pagestyle{plain} 
\setcounter{page}{1}
\pagenumbering{arabic}


\section{Introduction}
\label{sec:Introduction}
 
The underlying mechanisms governing hadron production in high-energy proton-proton~($pp$) collisions at the LHC are a collection of perturbative and nonperturbative partonic processes occurring at different momentum-transfer scales. The production of hadrons with large momentum component transverse to the beam, \pt, is dominated by hard parton scatterings, where only two partons with large momentum transfers are involved. Theoretical predictions can be made in the framework of perturbative Quantum ChromoDynamics~(QCD) by using the factorisation theorem to disentangle parton distribution functions, QCD matrix elements and the fragmentation~\cite{Collins:1989gx}. 
Hadrons produced instead at $\pt\lesssim 2\gevc$ originate predominantly from soft scatterings with momentum transfers of the order of $\Lambda_\mathrm{QCD}$ and below~\cite{Mueller:1970,Gross:1973,Christiansen_2025}. Effective field theories and QCD-inspired phenomenological models are implemented in event generators and tuned to accurately reproduce experimental measurements~\cite{BUCKLEY2011145,Sjostrand:2022oyu}.
This is crucial in ground-based cosmic-ray observatories, which rely on these phenomenological models to infer information on a primary cosmic ray from the hadronic shower it initiates when interacting with Earth's atmosphere.
In particular, an important quantity to infer the mass of the primary particle is the number of muons produced at the end of the hadronic cascade~\cite{Dembinski:2017kpa}. This is currently underestimated by common event generators in astroparticle physics, an observation referred to as the Muon Puzzle~\cite{muon_puzzle_at_LHC,global_tuning}.

While models based on string fragmentation~\cite{Andersson:1983ia,Bierlich:2022pfr,Pierog:2009zt,Pierog:2013ria,Ostapchenko:2010vb,Ostapchenko:2013pia,Lykasov:1995tu,Ahn:2009wx,Riehn:2019jet} are widely used to describe the hadronisation process, cluster fragmentation models~\cite{Bahr:2008pv,Bellm:2015jjp,Bellm:2019zci} provide an alternative approach. The hadronisation predominantly produces light $u,d$ and $s$ quarks, with the $s$-quark production suppressed with respect to that of $u$ and $d$ quarks through the $s$-quark suppression parameter~\cite{Bierlich:2022pfr}.
The main mechanism of baryon and antibaryon production proceeds through the production of diquark-antidiquark ($qq$-\quarkbar\quarkbar) pairs, which can be described with the diquark suppression parameter~\cite{Bierlich:2022pfr,Riehn:2019jet}.
However, contributions from the fragmentation process of the proton beam remnants may play a role in baryon production and the baryon number transfer~\cite{Andersson:1983ia,Bierlich:2022pfr,Pierog:2009zt,Pierog:2013ria,Ostapchenko:2010vb,Ostapchenko:2013pia,Lykasov:1995tu,Ahn:2009wx,Riehn:2019jet}. A measurement of antibaryon-to-meson suppression, such as the production cross-section ratio
\begin{equation*}
	R(\Lbar/\KS)=\frac{\sigma(pp\to\Lbar X)}{\sigma(pp\to\KS X)},
\end{equation*}
where $X$ represents any additional set of particles, is a powerful probe of antibaryon production independent of the hadronisation of the beam remnants due to the absence of a net strangeness in the colliding protons.
By performing differential measurements across different centre-of-mass energies, momentum and rapidity spectra, the information gain can be maximised. 
The LHCb experiment is particularly sensitive to these processes over a broad momentum spectrum at high rapidity thanks to a full instrumentation of the detector in the forward region that is unique among the LHC experiments. 

Differential measurements of $R(\Lbar/\KS)$ in rapidity $y$ and \pt have been carried out previously at the LHCb experiment in $pp$ collisions at centre-of-mass energies of \mbox{$\sqrt{s} = 0.9,\ 7$ and 13\tev}~\cite{LHCb-PAPER-2011-005,LHCb-PAPER-2026-031-title}.
In addition, the ALICE collaboration measured $R[(\Lz + \Lbar)/2\KS]$ in $pp$ collisions at $\sqrt{s}=0.9,\ 7,$ and 13\tev as a function of \pt~\cite{ALICE:2010vtz, ALICE_strange_hadrons_pp_7_13TeV, ALICE_multistrange_multiplicity_13TeV}, as well as the CMS collaboration at $\sqrt{s}=$ 0.9 and 7\tev differentially in $y$ and \pt~\cite{CMS:2011jlm}.
Furthermore, the STAR collaboration at RHIC-BNL has reported an integrated value of $R(\Lbar/\KS)$ using $pp$ collisions at $\sqrt{s}=200\gev$~\cite{STAR}.
In this paper, the $R(\Lbar/\KS)$ ratio is measured in intervals of $y$ and \pt of the \Lbar and \KS hadron, which are collectively identified as \Vzero hadrons, for the first time in $pp$ collisions at $\sqrt{s}=13.6\tev$ using data acquired with the LHCb upgraded detector in 2024. Only promptly produced  \Vzero hadrons are considered, \ie only those originating directly from the primary $pp$ interaction or from strong and electromagnetic decays of resonances with lifetimes smaller than $10^{-16}\,\si{\second}$. Strange hadrons are produced copiously at the LHC, with more than one \Lbar and \KS hadron per $pp$ collision in the geometrical acceptance of the LHCb detector on average~\cite{AlvesJunior:2018ldo}. As a result, large samples of both hadrons can be collected in short data-taking periods. This analysis uses data corresponding to an integrated luminosity of $\SI{64}{\pico\barn^{-1}}$.

\section{Detector and simulation}
\label{sec:Detector}
The \lhcb Run~3 detector~\cite{LHCb-DP-2022-002} is a single-arm forward spectrometer covering the pseudorapidity range $2 < \eta < 5$, designed for the study of particles containing \bquark or \cquark quarks.
Installed prior to the start of Run~3 data taking in 2022, it represents a major upgrade of the \lhcb detector. The new configuration
was designed to match the performance of the Run~1--2 detector~\cite{LHCb-DP-2014-002,LHCb-DP-2008-001}, while allowing it to operate 
at approximately five times the instantaneous luminosity in $pp$ collisions. 
The high-precision tracking system consists of a silicon-pixel vertex detector surrounding the $pp$ interaction region (VELO), a large-area silicon-strip detector (UT) located upstream of a dipole magnet with a bending power of about $4\,{\rm T\,m}$, and three stations of scintillating-fibre detectors (SciFi) downstream of the magnet.
Tracks reconstructed in both the VELO and the SciFi, with the possible addition of intermediate hits in the UT, are referred to as \textit{long} tracks. 
Different types of charged hadrons are distinguished using information from two ring-imaging Cherenkov detectors, equipped with photon detection systems.
Photons, electrons and hadrons are identified by a calorimeter system consisting of electromagnetic and hadronic calorimeters. Muons are identified by a system composed of alternating layers of iron and multiwire proportional
chambers~\cite{LHCb-DP-2025-007}.
The readout of all LHCb subsystems at the bunch-crossing frequency and the subsequent processing of the data in an all-software trigger~\cite{LHCb-TDR-016, LHCb-TDR-017, LHCb-TDR-018, LHCb-TDR-021} is a central feature of the upgraded detector, enabling the reconstruction and selection of events in real time.

The trigger system is implemented in two stages: a GPU-based inclusive stage (HLT1) focused primarily on charged particle reconstruction, which reduces the data volume by roughly a factor of 30~\cite{LHCb-TDR-021}, followed by a CPU-based stage (HLT2), which performs the physics-analysis-quality reconstruction and selection of physics signatures~\cite{LHCb-TDR-016}.
In between these two stages, an alignment and calibration of the detector is performed in near real-time and their results are used in the trigger~\cite{LHCb-PROC-2015-011}.
Triggered data further undergo a centralised, offline processing step to deliver physics-analysis-ready data across the entire \lhcb physics programme~\cite{Sprucing, FunTuple}.

Simulated events are used to optimise selection requirements and determine reconstruction and selection efficiencies. In the simulation, $pp$ collisions are generated using \pythia~\cite{Sjostrand:2007gs} with a specific \lhcb configuration~\cite{LHCb-PROC-2010-056}. Decays of unstable particles are described by \evtgen~\cite{Lange:2001uf}, in which final-state radiation is generated using \photos~\cite{davidson2015photos}. The interaction of the generated particles with the detector, and its response, are implemented using the \geant toolkit~\cite{Agostinelli:2002hh} as described in Ref.~\cite{Allison:2006ve}.

\section{Analysis strategy}
\label{sec:sel}

The production cross-section ratio $R(\Lbar/\KS)$ is measured using \decay{\KS}{\pip\pim} and \decay{\Lbar}{\antiproton\pip} decays, and is calculated as

\begin{equation}
    R(\Lbar/\KS)=\frac{N(\Lbar\to\antiproton\pip)}{N(\KS\to\pip\pim)}\frac{\varepsilon_{\KS\to\pip\pim}}{\varepsilon_{\Lbar\to\antiproton\pip}}\frac{\mathcal{B}(\KS\to\pip\pim)}{\mathcal{B}(\Lbar\to\antiproton\pip)},\label{eq:ratio}
\end{equation}
where the branching fractions $\mathcal{B}$ of the \Lbar and \KS decays are external inputs taken from the world average~\cite{PDG2026}, the efficiencies, $\varepsilon$, of the candidate reconstruction and selection are determined from simulation, and the signal yields $N$ are determined from invariant-mass fits to the candidates selected in data.

For the measurement described in this paper, data are acquired with an unbiased trigger, 
where randomly selected bunch crossings are kept. Moreover, signal candidates are selected using only kinematic and topological quantities, such that the measurement is insensitive to potential biases from a trigger selection and from effects related to the particle identification system.

\begin{figure}[tb]
\centering
\includegraphics[width=0.54\textwidth]{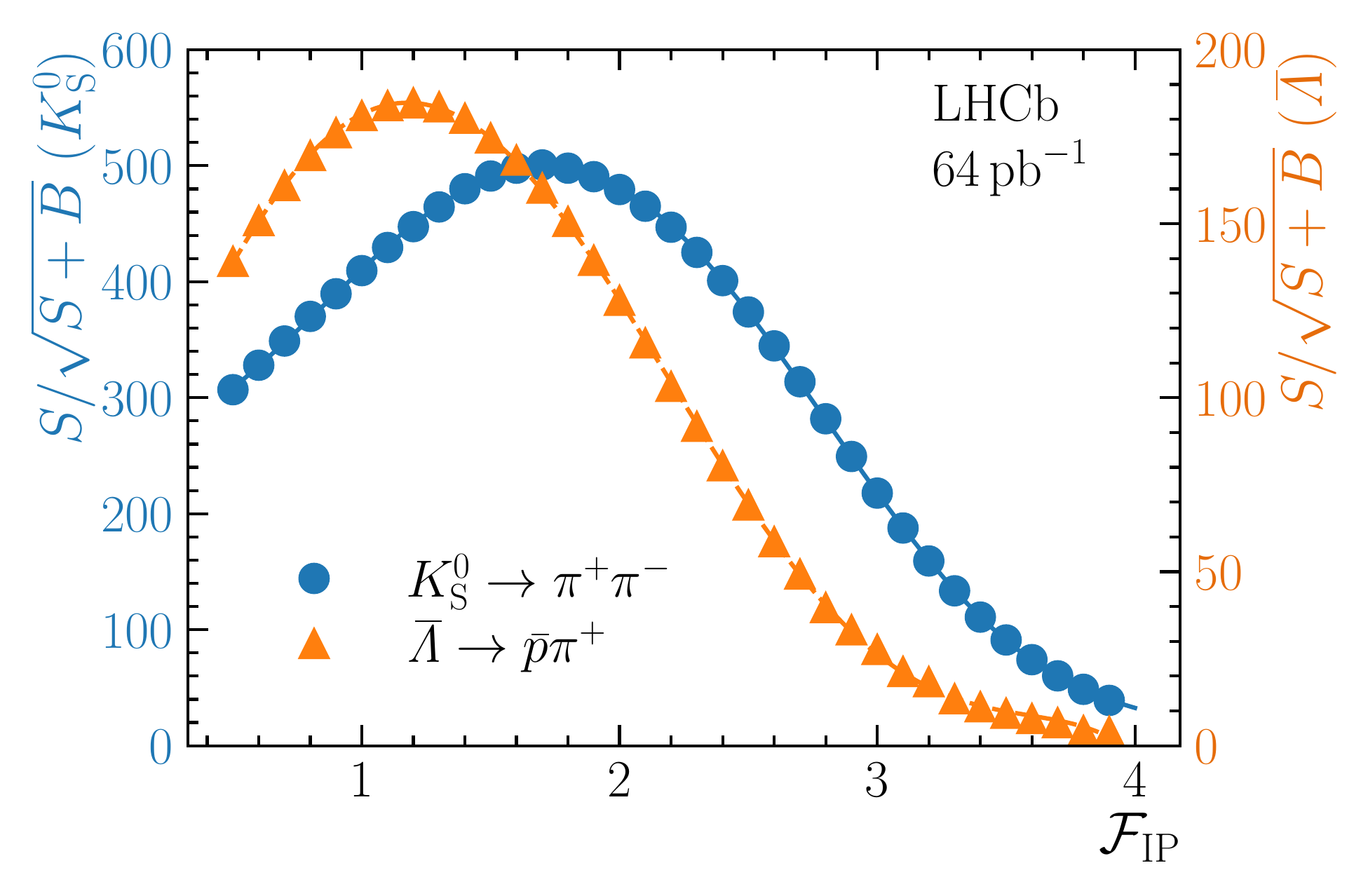}
\includegraphics[width=0.44\textwidth]{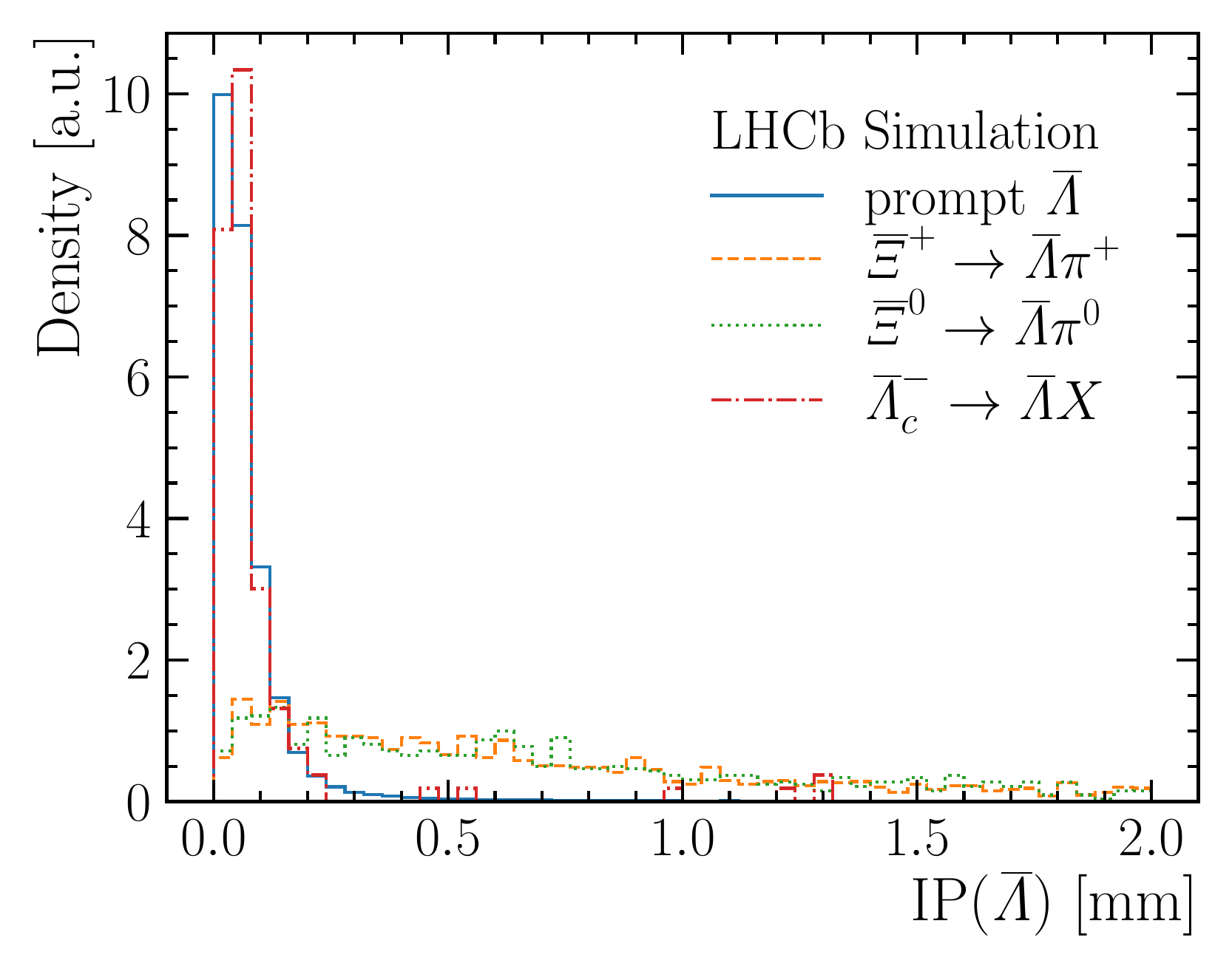} 
\caption{(Left) Scan of the $\mathcal{F}_\mathrm{IP}$ discriminant for \decay{\KS}{\pip\pim} and \decay{\Lbar}{\antiproton\pip} decays to find the optimal selection value. (Right) IP distributions of prompt and secondary \Lbar decays in simulation.}
\label{fig:IP_cut}
\end{figure}

Candidates of \KS and \Lbar hadrons are reconstructed from two oppositely charged tracks with mass hypotheses consistent with being pions or antiprotons that are combined to form a common vertex, required to lie within the VELO detector.
All reconstructed \Vzero candidates are required to have an invariant mass within $\pm50\mevcc$ of their respective known values~\cite{PDG2026}, and to lie in the kinematic fiducial region of $0.15 < \pt < 2.5\gevc$ and $2.0 < y < 4.5$.
The impact parameters (IPs), defined as the minimum distance between a particle’s trajectory and a primary $pp$ collision vertex, of the \Vzero candidate and its decay products are used to discriminate signal candidates against random track combinations. They are combined into a Fisher discriminant $\mathcal{F}_\mathrm{IP}$ as 

\begin{equation}
\mathcal{F}_\mathrm{IP}(V^0) = \log_{10}(\mathrm{IP}(h^+))+\log_{10}(\mathrm{IP}(h^{(\prime)-}))-\log_{10}(\mathrm{IP}(V^0)),
\end{equation}

\noindent
where $h^{+/(\prime)-}$ are the final-state particles of the $V^0 \to h^+ h^{(\prime)-}$ decay. Including the IP of the \Vzero itself ensures that the track combination points to a primary $pp$ interaction vertex and suppresses contributions from \Vzero particles produced in weak decays of hyperons ($\PXi^{-,\, 0}, \POmega^-$), charm and beauty hadrons as well as in hadronic interactions of charged particles with the LHCb detector material.  
Optimal $\mathcal{F}_\mathrm{IP}$ selection working points are determined independently for the \KS and \Lbar candidates by maximising the signal figure of merit $S/\sqrt{S+B}$, where $S$ and $B$ correspond to the signal and background yields obtained from fits to the invariant-mass distributions in data. A scan of the figure of merit as a function of $\mathcal{F}_\mathrm{IP}$ for \KS and \Lbar hadrons is shown in Fig.~\ref{fig:IP_cut}. The optimal selections are found to be $\mathcal{F}_\mathrm{IP}(\KS)>1.7$ and $\mathcal{F}_\mathrm{IP}(\Lbar)>1.2$. Decays of the \KS meson in the \Lbar selected sample are removed by assigning the antiproton candidate the pion mass and vetoing all candidates in the resulting two-pion invariant mass in the range of $\pm15\mevcc$ around the known \KS mass~\cite{PDG2026}. In a similar way, \Lbar and \Lz decays are removed from the \KS candidate sample with a veto mass window of $\pm 10\mevcc$ around the known \Lbar mass~\cite{PDG2026}. Backgrounds from nonprompt \Lbar coming from hyperon decays, as illustrated in Fig.~\ref{fig:IP_cut}, are further suppressed by applying an additional $\text{IP}(\Lbar)<\SI{0.15}{\milli\meter}$ requirement. This requirement reduces the nonprompt fraction of \Lbar decays to $\mathcal{O}(1\%)$.
The value is chosen from a scan in steps of 0.005\mm such that the decrease of the figure of merit is 1\%. 
After the full selection is applied, residual \Vzero nonprompt contributions are estimated from simulation for both \Lbar and \KS candidates, and subtracted from the measured signal yields.
Multiple candidates originating from the same $pp$ collision are permitted in the selection. A potential effect of this choice has been probed and is found to be negligible.

\begin{figure}[b]
    \centering
        \includegraphics[width=0.49\textwidth]{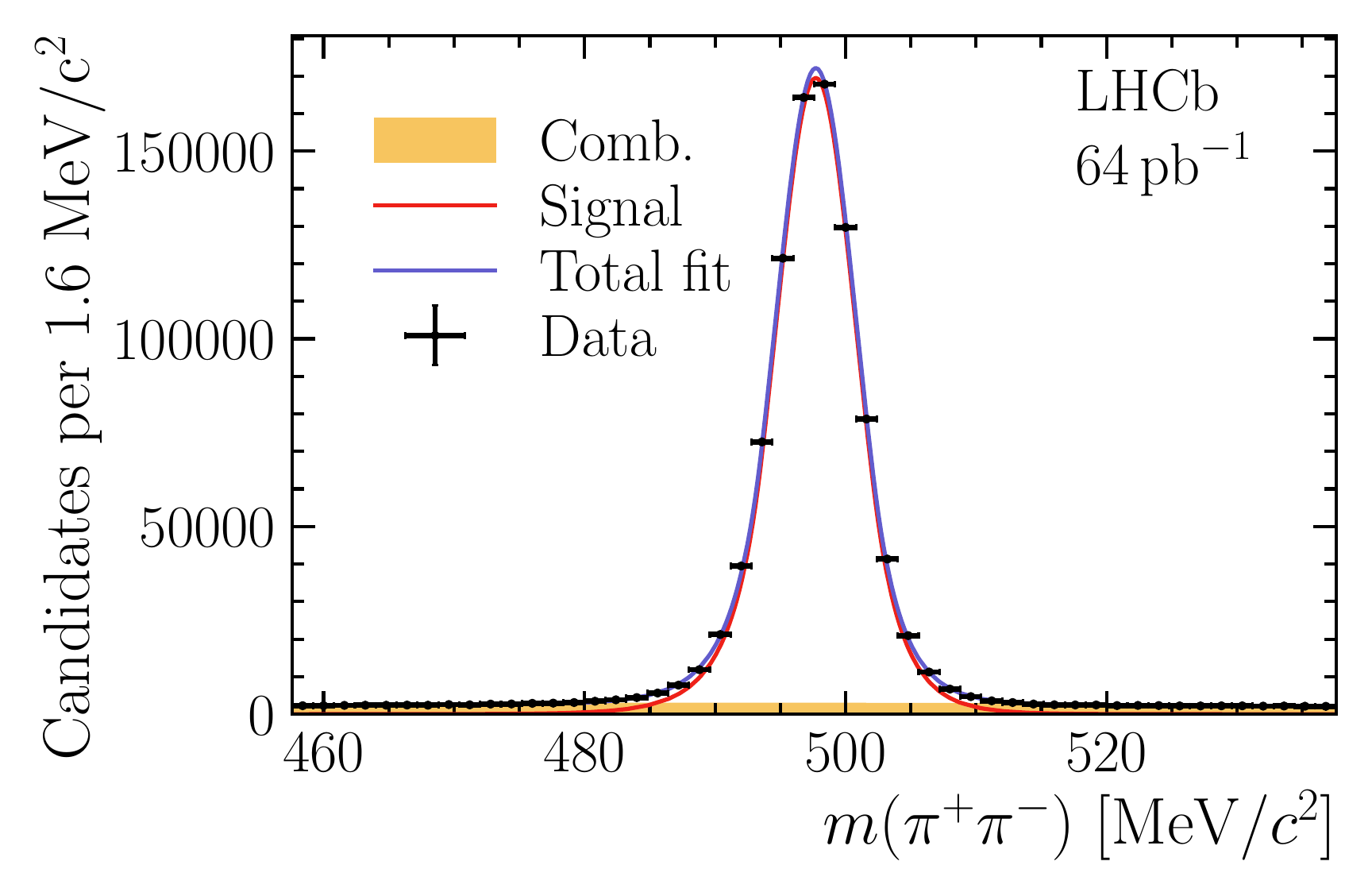}
        \includegraphics[width=0.49\textwidth]{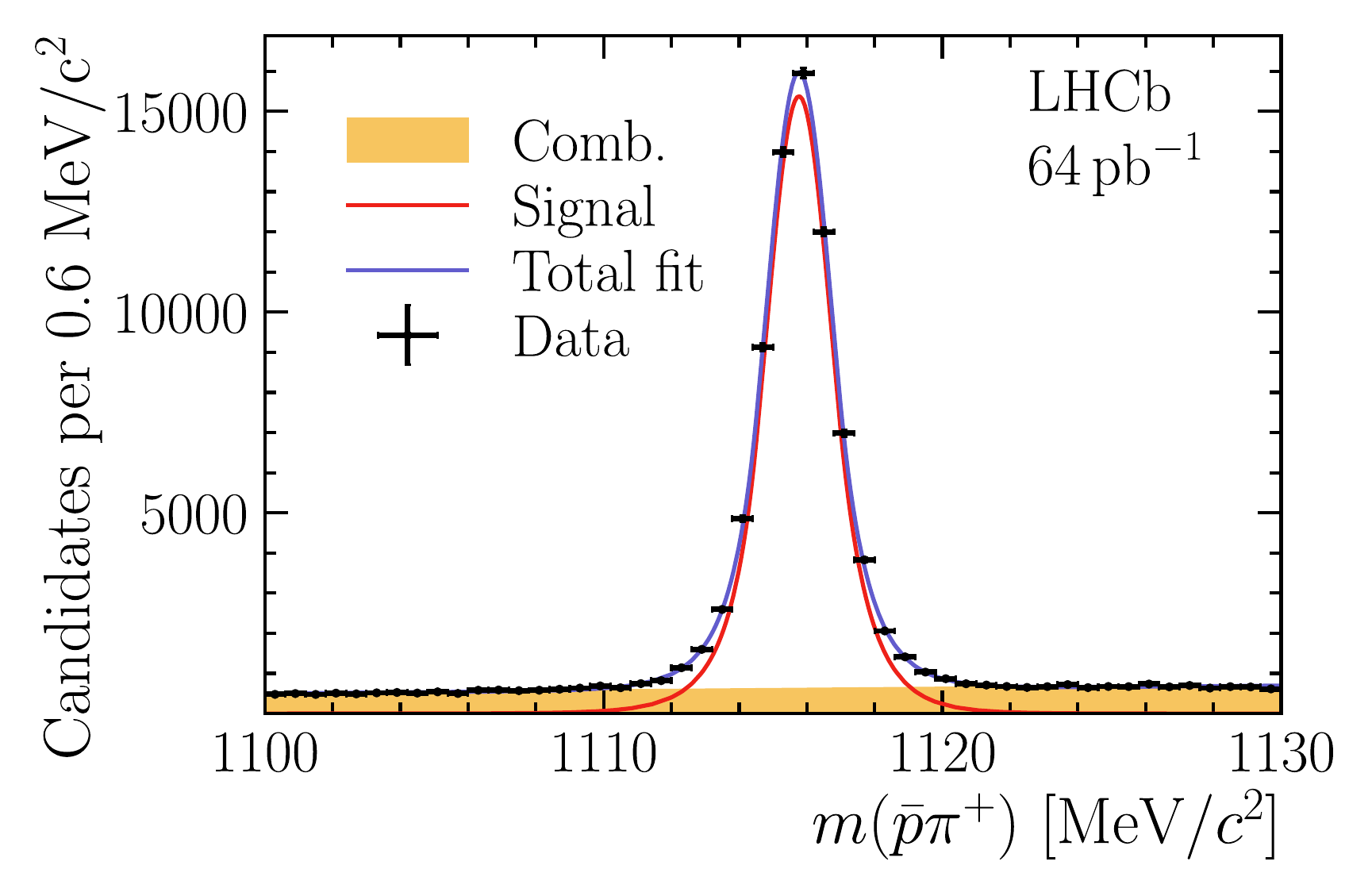}
    \caption{Invariant-mass distributions of the (left) \KS and (right) \Lbar selected candidates in the total data sample, with the fit results overlaid.}
    \label{fig:integrated_yield_Ks_L0}
\end{figure}

Signal yields of the \Vzero hadrons are obtained by performing unbinned extended maximum-likelihood fits to the invariant-mass distributions. The signal is parametrised with a double-sided Crystal Ball~\cite{Skwarnicki:1986xj} function (DSCB), while the background is modelled with a linear function. The DSCB parameters are initially obtained from fits to the simulation, and then fixed in the fit to the data, except for the mean and widths that are free to float to account for differences between simulation and data in the tracking detector alignment and mass resolution. 
Figure~\ref{fig:integrated_yield_Ks_L0} shows the \KS and \Lbar invariant-mass distributions for the total data sample, together with the overlaid fit results giving signal yields of $N(\KS\to\pip\pim) = 870\,900\pm1100$ and $N(\Lbar\to\antiproton\pip) = 69\,930 \pm 330$. The dataset is then split into \pt and $y$ intervals in the ranges $0.15 < \pt < 2.5\gevc$ and $2.0 < y < 4.5$. The invariant-mass fits are repeated using the same signal and background parametrisations in each of the kinematics intervals. 

\section{Estimation and calibration of efficiencies}
\label{sec:efficiencies}
The imperfect modelling of the production kinematics, detector alignment and reconstruction efficiency in the simulation introduces differences with respect to the data that can bias the final results. While the measurement of a ratio of production cross-sections has the advantage to be largely insensitive to many of these effects, discrepancies between data and simulation that affect \KS and \Lbar candidates in different ways need to be accounted for. First, a kinematic weighting in \Vzero \pt and $\eta$ is performed to match the simulated distributions to those seen in background-subtracted data~\cite{Pivk:2004ty}. The weighting is performed using the gradient-boosted weighting algorithm \texttt{GBReweighter} from the \texttt{hep\_ml} package~\cite{hep_ml}. After the kinematic weighting, residual discrepancies are observed between simulation and data in the \Vzero IP distributions, with a significantly larger deviation for \Lbar than for \KS candidates. Since the IPs of the \Vzero and its decay products serve as the main selection variables, achieving excellent agreement between data and simulation is crucial. To account for the observed discrepancy, likely caused by the imperfect modelling of the uncertainties in the reconstructed VELO hits, the IP values are scaled by a factor determined by minimising the \chisq value between the scaled distribution from simulation and that from background-subtracted data. This procedure yields scale factors of 0.98 for the IP of \KS candidates and 1.26 for the IP of \Lbar candidates, while for the IP of the antiproton in the \Lbar final state a scale factor of 0.99 is found and for final-state pions the description of the IP in data is already very good and cannot be enhanced via scaling.

A tag-and-probe approach is used to estimate differences in the track reconstruction efficiencies between data and simulation.
For the SciFi track reconstruction efficiency for long tracks, $\varepsilon_\mathrm{trk}^\mathrm{long|SciFi}$, \Vzero candidates are reconstructed by combining a long track~(tag), \ie that leaves detector hits in the VELO and SciFi with optional hits in the UT, with another track that is only reconstructed with information from the VELO detector~(probe).
The momentum of the probe track is inferred from the knowledge of the decay vertex of the \Vzero candidate and its associated primary vertex, as well as the VELO track direction and momentum of the tag track.
A fit to the invariant-mass distribution of the \Vzero candidates in this data sample is performed to determine the signal yield $N^\text{tag+VELO}$ and compare it to the signal yield $N^\text{tag+VELO}_\mathrm{matched}$ when both tracks of the \Vzero candidate are reconstructed with the full tracking system. 
The same procedure is repeated to obtain the VELO track reconstruction efficiency, $\varepsilon_\mathrm{trk}^\text{VELO}$, by using a track only reconstructed with the SciFi and UT detectors as probe.
Here, the signal yield $N^\text{tag+(SciFi+UT)}$ is compared against $N^\text{tag+(SciFi+UT)}_\mathrm{matched}$
The total track reconstruction efficiency is obtained as the product of efficiencies
\begin{equation}
\varepsilon_\mathrm{trk} = \varepsilon_\mathrm{trk}^\text{VELO}\varepsilon_\mathrm{trk}^\mathrm{long|SciFi} = \frac{N^\text{tag+(SciFi+UT)}_\mathrm{matched}}{N^\text{tag+(SciFi+UT)}}\frac{N^\text{tag+VELO}_\mathrm{matched}}{N^\text{tag+VELO}}.
\end{equation}
\noindent
The derivation of the track reconstruction efficiency is done separately for \pip and \pim tracks from \KS and $\bar{p}$ tracks from \Lbar candidates and performed in intervals of the final-state particle momentum, $p$, $\eta$ and the polar angle, $\varphi$, in both data and simulation. The ratios of efficiencies between data and simulation are taken as per-track correction weights for the track reconstruction efficiency, and used to correct the efficiency estimation from simulation.
To ensure sufficient data for the three-dimensional correction map, additional data from the same period corresponding to an integrated luminosity of \SI{529}{\pico\barn^{-1}} are used in this study. These are otherwise not included into the analysis as the statistical uncertainty of the results is subdominant.

\begin{figure}[tb]
    \centering
        \includegraphics[width=0.49\textwidth]{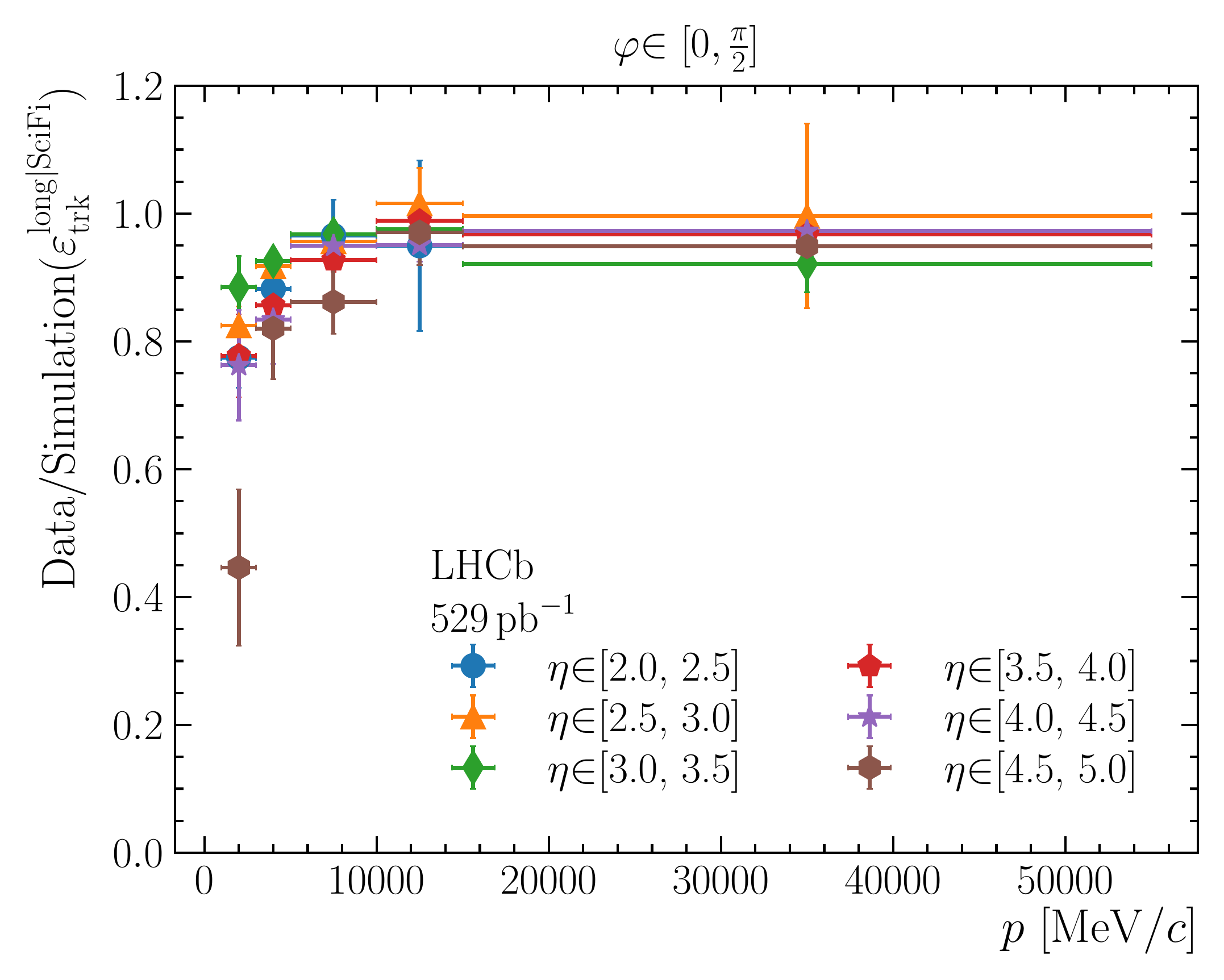}
        \includegraphics[width=0.49\textwidth]{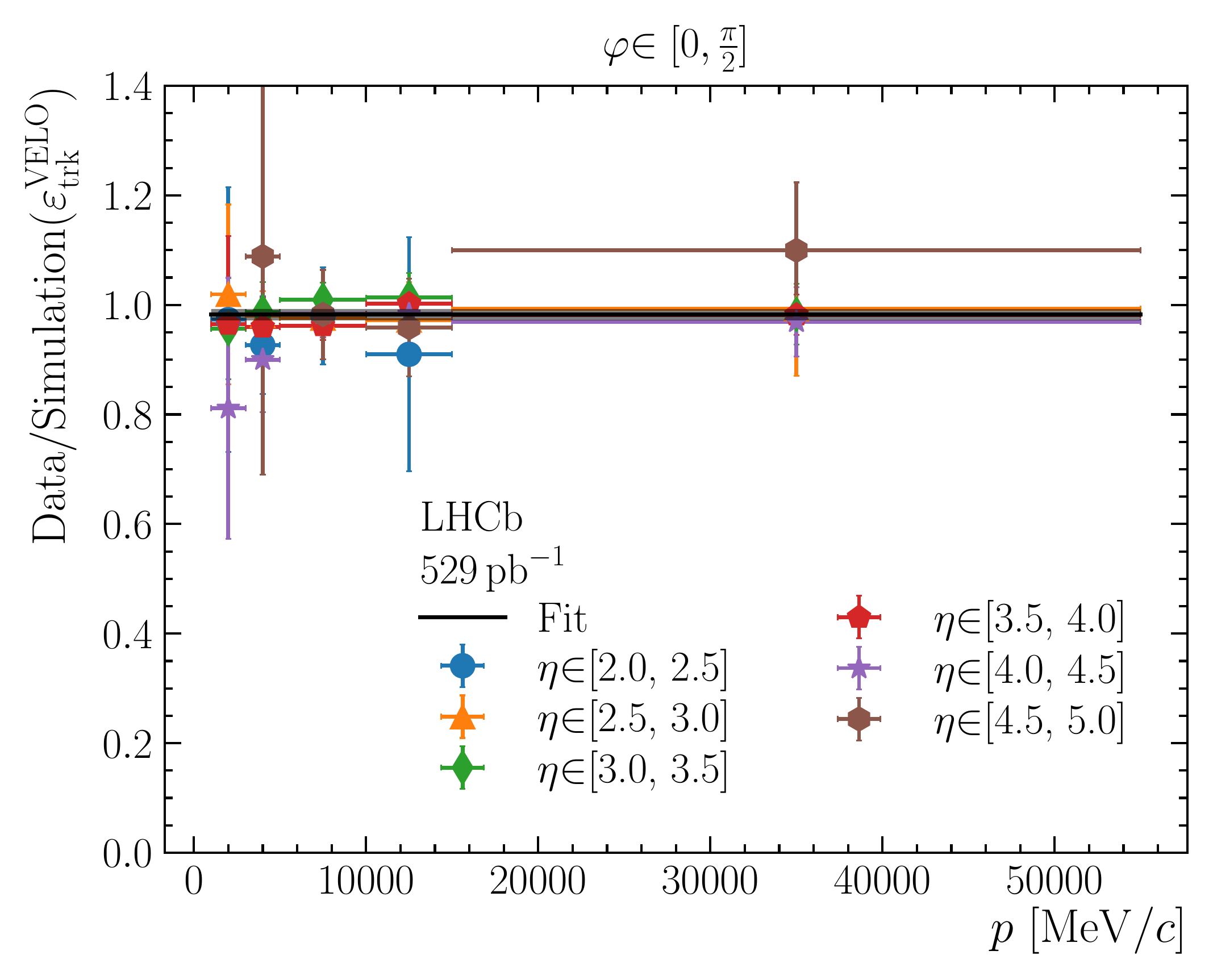}
    \caption{Ratio of the \pim (left) SciFi efficiency for long tracks and (right) VELO track reconstruction efficiencies in data and simulation as a function of \textit{p} and $\eta$ in the region $\varphi \in [0,\pi/2]$. 
    For the VELO efficiency a linear fit is performed to probe the average offset from unity. } 
    \label{fig:SciFi_VELO_track_eff}
\end{figure}

The SciFi track reconstruction efficiency ratios for long tracks in data and simulation are shown in Fig.~\ref{fig:SciFi_VELO_track_eff} for \pim mesons originating from \KS decays as a function of $p$ and $\eta$ in one of the probed $\varphi$ regions.
The corrections are generally in the range from $0.8$ to unity with larger values for high-$\eta$ tracks. No significant differences between particle species or charges are observed.
Overall, the obtained correction weights show a good description of the detector performance in the central region, while larger deviations are observed to the edges of detector acceptance.
The VELO tracking efficiency ratios for \pim in the same geometric region are also shown in Fig.~\ref{fig:SciFi_VELO_track_eff}. 
The ratios show no significant deviation from unity, therefore no correction is applied. A systematic uncertainty of $\SI{2}{\percent}$ is assigned for this choice, as discussed in Sec.~\ref{sec:syst}. By construction, the approach accounts for potential mismodelling of the hadronic interaction of the decay products with the traversed detector material.
As these interactions can lead to not reconstructing the \Vzero candidate, it would be reflected in the tracking efficiency ratio. This covers uncertainties on the total detector material budget and the hadronic interaction cross-sections.

The \Vzero efficiencies are factorised into a term related to the \Vzero candidate reconstruction and one concerning its selection. Reconstruction efficiencies are determined per kinematic interval by summing the product of kinematic and track reconstruction weights for each reconstructed candidate in the simulation and divided by the sum of kinematic weights of generated events. The selection efficiency is obtained by dividing the number of reconstructed and selected \Vzero candidates by the total number of reconstructed \Vzero candidates, applying both the track reconstruction correction weights as well as the kinematic weights and the IP scaling.

When using simulation, a matching between reconstructed and generated candidates is applied to determine the truth-level information of the particles.
To correctly identify a generated \Vzero hadron in the reconstructed simulation, both reconstructed final-state tracks and the reconstructed \Vzero candidate have to be matched to the corresponding generated particle, which is required to originate from the primary $pp$ interaction or from strong or electromagnetic decays.
Due to the imperfect efficiency of this matching procedure, a fraction of up to $4\%$ of the generated and reconstructed signal, where the matching of only one of the reconstructed final-state tracks to its generated particle succeeds, is lost.
To account for this fraction in the efficiency determination, a control sample is obtained from candidates that could not be identified as generated \Vzero hadrons by requiring one of the tracks in the final state to be matched to an antiproton or pion that has a \Vzero hadron as ancestor.
An invariant-mass fit is performed in the control sample and the measured yield $N_\mathrm{sel, \,unmatched}$ of unmatched particles is used to apply a correction determined as $1 + N_{\mathrm{sel, \,unmatched}}/{N_{\mathrm{sel}}}$ to the total efficiency, where $N_{\mathrm{sel}}$ is the number of selected candidates in simulation that could be matched to a generated particle.

\section{Systematic uncertainties}
\label{sec:syst}
Systematic uncertainties on $R(\Lbar/\KS)$ can be divided into those affecting the extraction of the observed signal yields, those related to the estimation of the reconstruction and selection efficiencies, and  external inputs. A summary of the systematic uncertainties is given in Table~\ref{tab:systematics}.

The systematic uncertainty on the extracted signal yields due to the choice of the parametrisations to model the signal and background contributions is evaluated by replacing the DSCB as signal model with a Student's t-distribution. Similarly, an alternative background model is considered by replacing the linear function with a second-order polynomial. The difference in the signal yield from both strategies is added in quadrature and taken as total systematic uncertainty for the fit model.

A potential bias due to wrongly estimated residual nonprompt contributions could scale the signal yields and introduce kinematic-dependent differences between \KS and \Lbar. While the production cross-sections of charm and beauty hadrons and their decay rates to \Vzero particles have been measured to good precision, the production cross-sections of hyperons are known less precisely and could potentially be mismodelled in the simulation. At the same time, fits to the IP distributions indicate that the nonprompt component, contributing at the order of $\SI{1}{\percent}$, is modelled correctly. Therefore, a 50\% relative uncertainty on the residual nonprompt fractions, estimated from simulation, is assigned for each kinematic interval.

The systematic uncertainty on the correction to the SciFi efficiency for long tracks due to the limited amount of data is evaluated using pseudoexperiments. For each kinematic interval, 50 pseudoexperiments are randomly sampled from a Gaussian distribution centered on the nominal value of the correction weight, and the width corresponding to its uncertainty. The variation of the results is taken as systematic uncertainty. An additional systematic uncertainty is added to account for the variations in the correction weights to the VELO tracking efficiency. This uncertainty is determined via a constant fit to the track reconstruction correction weights in each $\varphi$ quadrant, as seen in Fig.~\ref{fig:SciFi_VELO_track_eff}. The largest deviation from unity among these fits is 2\% and is applied as a systematic uncertainty. An additional systematic uncertainty in the tracking efficiency arises from variations in the hit detection efficiency performance in one side of the SciFi detector during the 2024 data taking which affects the efficiency for the \pip candidate. Half of the difference on the tracking efficiency for the \pip candidate, which is bent towards an inefficient region by the dipole magnet of the detector, between the calibration data with and without this inefficiency is taken as a systematic uncertainty.
Furthermore, a systematic uncertainty due to the bin migration effect is assigned to the momentum inference method for the VELO probe tracks. The transfer matrix between inferred and matched long-track momentum is calculated and new tracking corrections are obtained by folding the momentum distribution with the transfer matrix. 
The difference between these and the nominal corrections are assigned as a systematic uncertainty.

A systematic effect in the efficiency calculation can arise due to residual data-simulation mismatches in the IP-based selection variables.
The cross-section ratios are calculated with and without the IP requirement on the \Lbar, which is used to suppress nonprompt background contributions. The ratio between these two results is calculated for each \pt and $y$ interval. A fit of a constant function is applied to this ratio and the relative shift is taken as a global systematic uncertainty for all bins. The uncertainty on the fit constant is applied as a local uncertainty to account for trends in the \pt and $y$ intervals.

Not only the charged decay products but also the \Vzero itself can interact with the material while traversing it and therefore can be lost for detection. Uncertainties in the total material budget as well as differences between simulation and data in the interaction cross-sections of the \Vzero candidates and decay products with the material can lead to biases in the reconstruction efficiencies. While the biases due to material interactions of the decay products are accounted for by the track reconstruction efficiency corrections, for the \Vzero a systematic uncertainty is evaluated separately. To measure the impact of this effect on the final result, the ratio between data and simulation of the \Vzero yield ratio is measured in intervals of the traversed material. The weighted variance of the ratio across these intervals is taken as systematic uncertainty.

The statistical uncertainty on the fitted yields in the control sample for the unmatched track correction is considered as a systematic uncertainty.
Finally, the uncertainty on the \Vzero decay branching fractions of \SI{0.8}{\percent} is taken into account as an external source of systematic uncertainty, where the measurements are taken from the world average~\cite{PDG2026}.

\begin{table}
    \centering
    \caption{Summary of the systematic uncertainties. Ranges are given for the variation of the systematic uncertainties across kinematic intervals.}
    \begin{tabular}{l c}
        \toprule
         Alternative mass-fit models & $0.4-5.7\%$ \\
         Nonprompt backgrounds & $0.8-1.6\%$ \\
         Reconstruction efficiency & $0.8-17.6\%$ \\
         Selection efficiency, global & $2.2\%$ \\
         Selection efficiency, local & $1.2\%$\\
         Material interactions & $3.0\%$\\
         Unmatched-track correction & $0.2-3.3\%$ \\
         \Vzero branching fraction  & $0.8\%$\\
         \midrule
         Total systematic uncertainty & $4.2-18.2\%$\\
         Statistical uncertainty & $0.8-5.4\%$\\
        \bottomrule
    \end{tabular}
    \label{tab:systematics}
\end{table}

\section{Results}
\label{sec:results}
\begin{figure}[tb]
        \centering
        \includegraphics[width=0.7\textwidth]{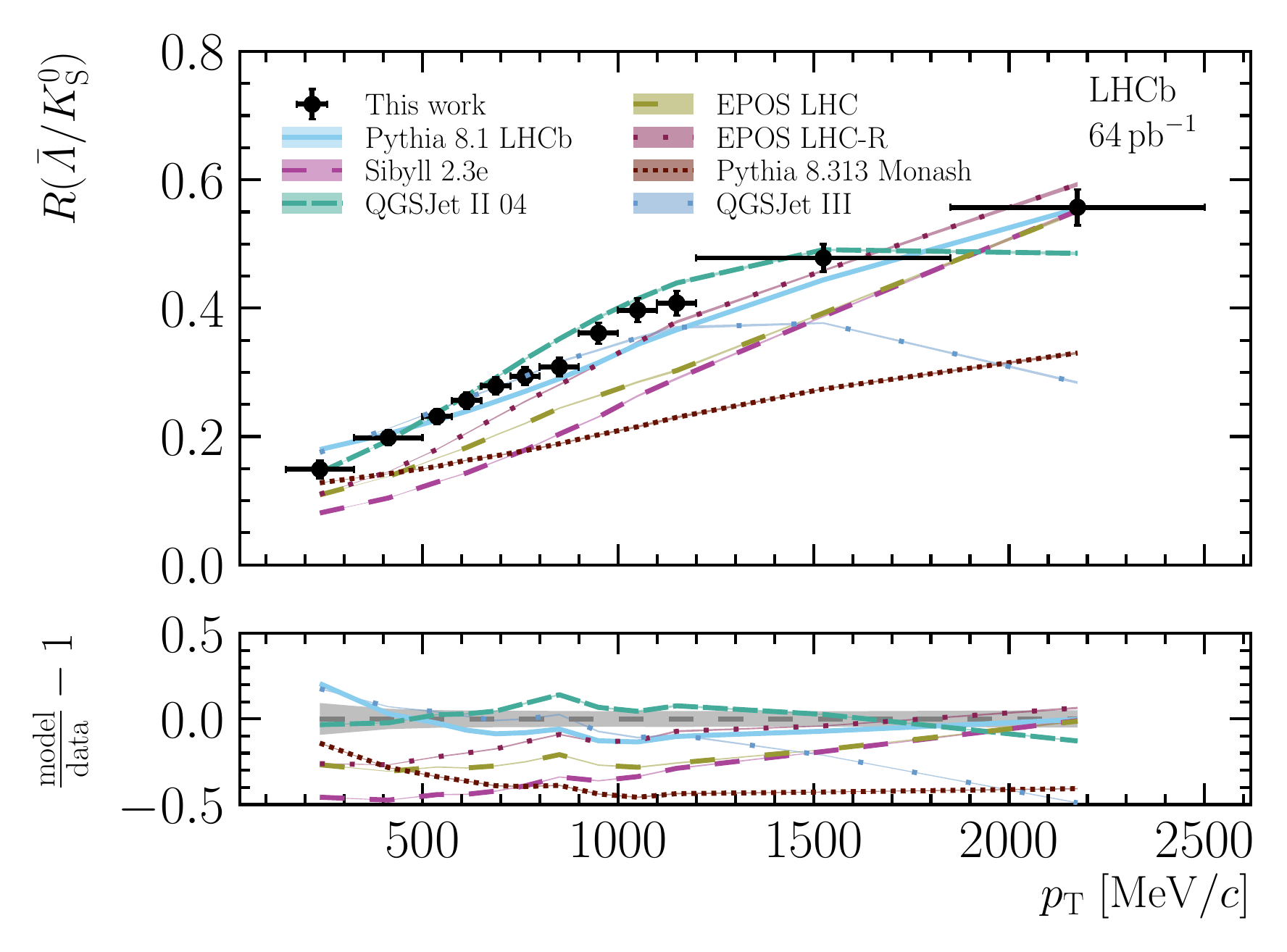}
        \includegraphics[width=0.7\textwidth]{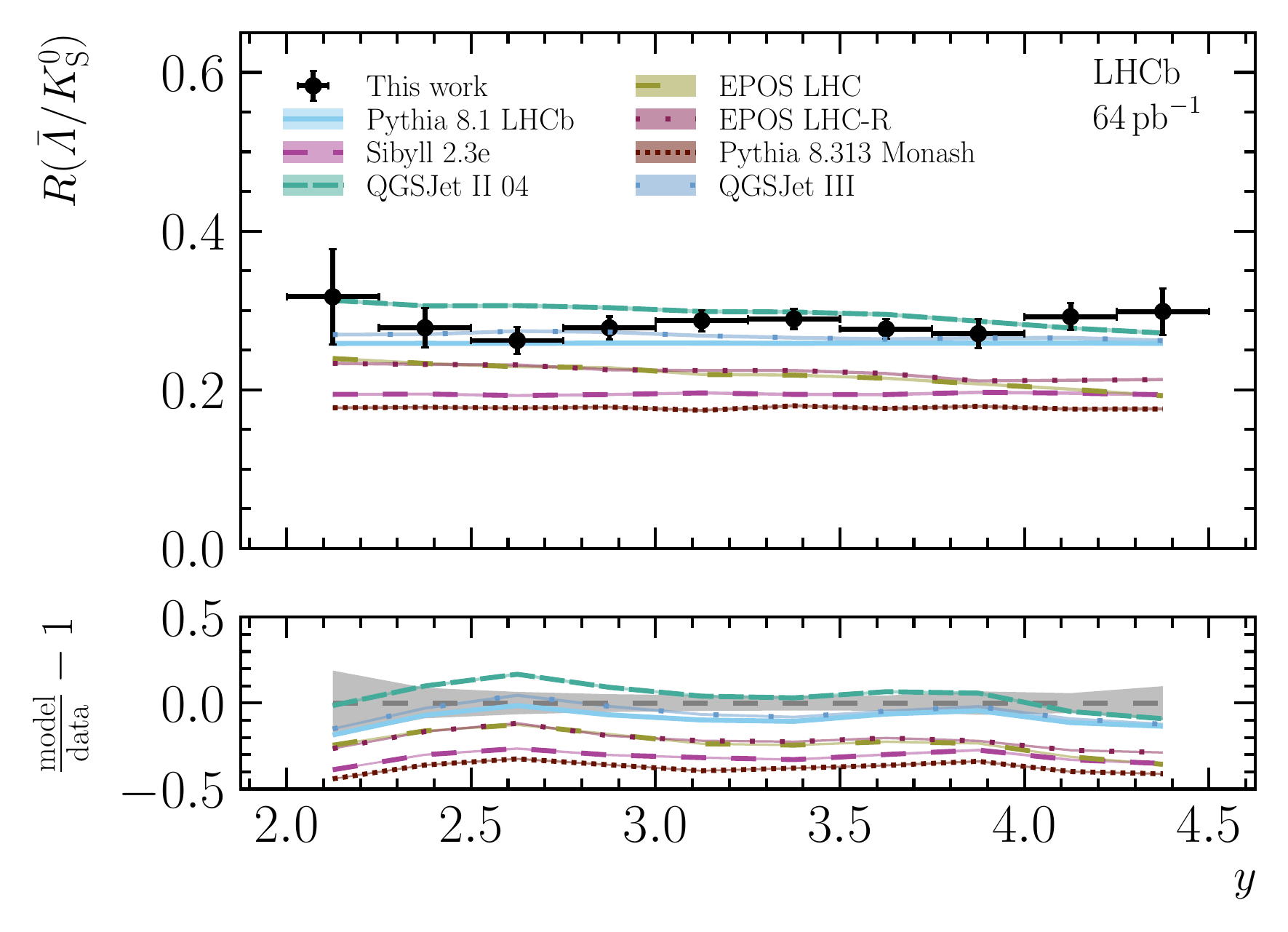}
    \caption{Prompt production cross-section ratios as a function of (top) \pt  and (bottom) $y$ compared to predictions from different generators. Coloured bands around the generator predictions indicate their statistical uncertainties, while the errorbars in the upper and the grey bands in the lower panels corresponds to the total uncertainty of this measurement.}
    \label{fig:result_generators}
\end{figure}

The measurement of the differential production cross-section ratios is shown in Fig.~\ref{fig:result_generators} and compared to predictions of commonly used event generators, produced using the \textsc{Chromo} software package~\cite{fedynitch2026chromo}. The correlation matrices of the measurements are provided in Appendix~\ref{app:correlation}. The results and correlation matrices are also made available on \texttt{HEPData}~\cite{hepdata}. 
The following event generators and tunes are considered: \pythia~8.1~LHCb~\cite{Sjostrand:2007gs,LHCb-PROC-2010-056}, \pythia~8.313~Monash~\cite{Skands:2014pea}, EPOS~LHC~\cite{Pierog:2013ria}, EPOS~LHC-R~\cite{Pierog:2025ixr}, Sibyll~2.3e~\cite{Ahn:2009wx,Riehn:2019jet}, QGSJet~II~04, and QGSJet~III~\cite{Ostapchenko:2010vb, Ostapchenko:2013pia}.
\pythia is a general-purpose event generator that is mostly used in collider physics. The other listed generators are optimised for describing particle production in high-energy collisions, including the interactions needed for cosmic-ray air showers.
While \pythia~8.313~Monash is mainly tuned using LEP, ATLAS, and CMS data, \pythia~8.1~LHCb is additionally tuned using the Run~1 \lhcb measurements of \Vzero production ratios and \KS production~\cite{LHCb-PAPER-2011-005}.
The EPOS generators describe the collision as a quantum-mechanical multiple-scattering problem, with energy-momentum conservation and detailed treatment of the subsequent hadronisation.
EPOS LHC has mainly been tuned to LEP data and early LHC measurements at 7\tev.
EPOS LHC-R is a recent development that has been tuned to LHC measurements at 13\tev, results obtained in lead-lead collisions with inputs from NA61.
Additionally, ``beam-particle color transparency" has been introduced in EPOS LHC-R, which reduces the strong interaction of the beam particles and helps modelling the hadronisation process more precisely.
Similarly to EPOS, QGSJET is a high-energy hadronic interaction model based on Gribov–Regge theory and multiple scattering.
It models cross-sections, secondary-particle production, diffraction, and nuclear interactions.
Sibyll is based on a QCD minijet model.

The \pt dependence of the $R(\Lbar/\KS)$ measurement is best described by QGSJet~II~04.
While also the LHCb tune \pythia~8.1 describes the data comparably well, the Monash tune vastly underestimates the ratio. EPOS LHC and Sibyll yield significantly lower values across most of the probed \pt range, however, their predictions approach the measured values at high \pt.
EPOS LHC-R underestimates the values at low to medium \pt, while at high \pt the predictions are comparable to those from the \pythia~8.1 LHCb tune.
QGSJet~III is also able to describe the data at low \pt before diverging towards lower values.

In the rapidity dependence of the cross-section ratio the predicted absolute values differ for Sibyll, EPOS~LHC and Pythia~8.313~Monash.
While some of these generators are able to describe the data much better than others, none of the probed event generators can describe the measured cross-section ratio of \Lbar and \KS accurately over the whole measurement space.
This highlights the significant impact of this measurement on the modelling of particle production in hadronic showers in the forward region, particularly for future tunes of event generators. 

\begin{figure}[tb]
    \centering
    \includegraphics[width=0.9\linewidth]{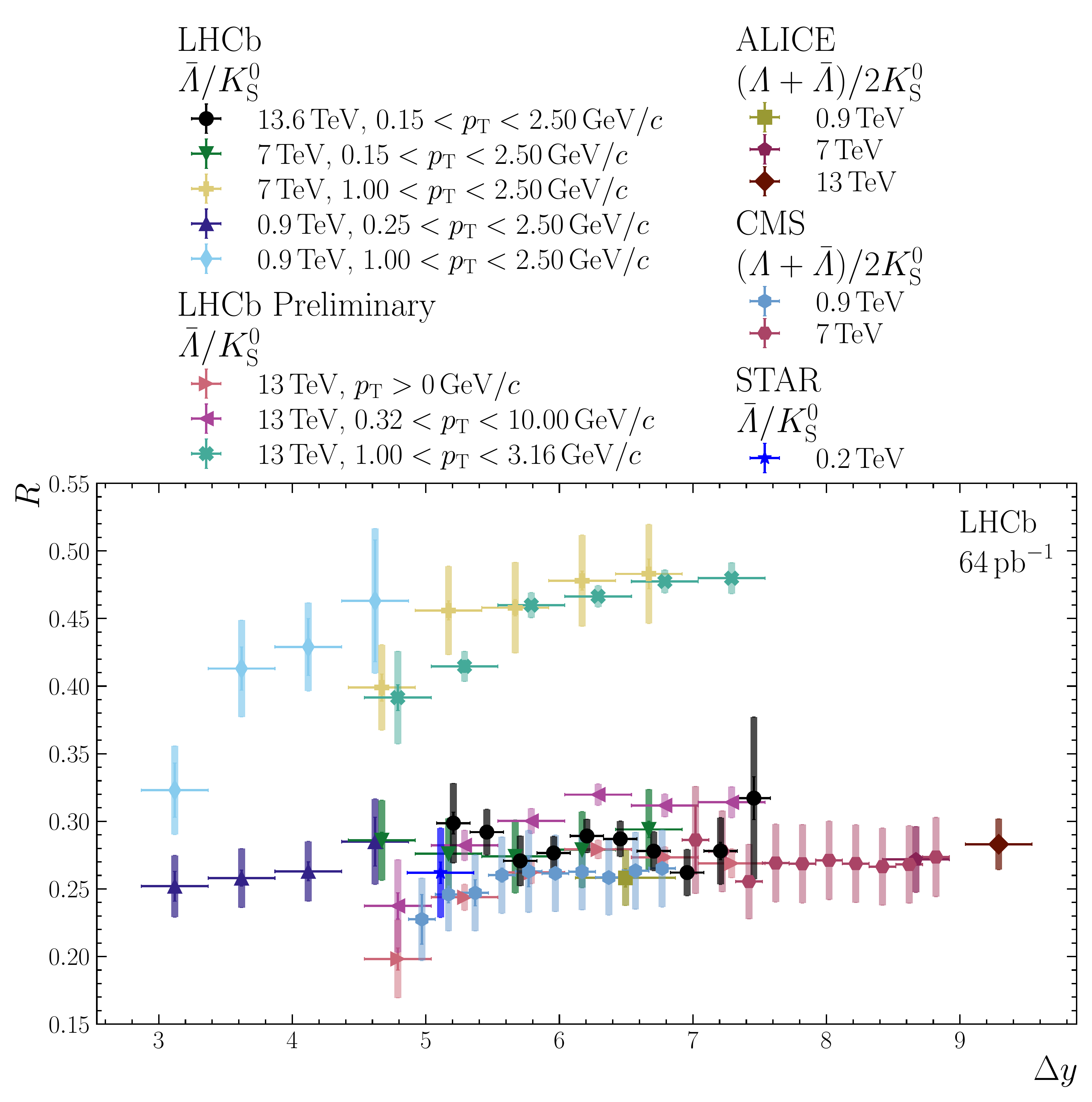}
    \caption{Prompt production cross-section ratios as a function of rapidity loss compared to previous \lhcb~\cite{LHCb-PAPER-2011-005,LHCb-PAPER-2026-031-title}, ALICE~\cite{ALICE:2010vtz, ALICE_strange_hadrons_pp_7_13TeV, ALICE_multistrange_multiplicity_13TeV}, CMS~\cite{CMS:2011jlm} and STAR~\cite{STAR} measurements in $pp$ collisions at different centre-of-mass energies.}
\label{fig:rapidity_loss}
\end{figure}

Figure~\ref{fig:rapidity_loss} compares the obtained results with current experimental determinations of $R(\Lbar/\KS)$ from previous \lhcb~\cite{LHCb-PAPER-2011-005,LHCb-PAPER-2026-031-title}, ALICE~\cite{ALICE:2010vtz, ALICE_strange_hadrons_pp_7_13TeV, ALICE_multistrange_multiplicity_13TeV}, CMS~\cite{CMS:2011jlm} and STAR~\cite{STAR} measurements as a function of the rapidity loss variable defined as $\Delta y \equiv y_\text{beam}-y$.
Here, $y_\text{beam}$ denotes the rapidity of the proton beam.
When the same \pt range is considered, good agreement is found between the first measurement at 13.6\tev presented in this paper and the previous measurements at $\sqrt{s}=0.9, 7$ as well as 13\tev. The ratio $R(\Lbar/\KS)$ shows no significant trend as a function of $\Delta y$. An offset is observed when comparing the result from this work with measurements at $\pt>1\gevc$ where the ratio $R(\Lbar/\KS)$ is significantly higher and exhibits an increasing trend with $\Delta y$. This can be explained by the \pt dependence, also shown in Fig.~\ref{fig:result_generators}.

\section{Conclusion}

The ratio of the prompt production cross-sections $R(\Lbar/\KS)$ as a function of \pt and $y$ is measured for the first time in $pp$ collisions at a centre-of-mass energy of $\sqrt{s}=13.6\tev$ using \SI{64}{\pico\barn^{-1}} of data recorded with the Run 3 LHCb detector in 2024.
Within the considered kinematic range of $0.15 < \pt < 2.5\gevc$ and $2.0 < y < 4.5$,
the measured cross-sections ratios are found to be in good agreement with previous measurements when compared as a function of the rapidity loss.
The results are further compared to the predictions of a set of event generators, where the best agreement is found with the QGSJet~II~04 model and the \pythia~8.1 LHCb tune. The results presented in this paper will enable future tunings of these generators and contribute to the understanding of fragmentation models for hadron production.
The agreement with previous measurements also provides an important validation of the understanding of the Run~3 LHCb detector, and in particular the determination of track reconstruction efficiencies with the fully replaced set of tracking detectors.

\section*{Acknowledgements}
%
%
\noindent We express our gratitude to our colleagues in the CERN
accelerator departments for the excellent performance of the LHC. We
thank the technical and administrative staff at the LHCb
institutes.
We acknowledge support from CERN and from the national agencies:
ARC (Australia);
CAPES, CNPq, FAPERJ and FINEP (Brazil); 
MOST and NSFC (China); 
CNRS/IN2P3 and CEA (France);  
BMFTR, DFG and MPG (Germany);
NKFIH (Hungary);              
INFN (Italy); 
NWO (Netherlands); 
MNiSW and NCN (Poland); 
MEC/IFA (Romania); 
MICIU and AEI (Spain);
SNSF and SER (Switzerland); 
NASU (Ukraine); 
STFC (United Kingdom); 
DOE NP and NSF (USA).
We acknowledge the computing resources that are provided by ARDC (Australia), 
CBPF (Brazil),
CERN, 
IHEP and LZU (China),
IN2P3 (France), 
KIT and DESY (Germany), 
INFN (Italy), 
SURF (Netherlands),
Polish WLCG (Poland),
IFIN-HH (Romania), 
PIC (Spain), CSCS (Switzerland), 
GridPP (United Kingdom),
and NSF (USA).  
We are indebted to the communities behind the multiple open-source
software packages on which we depend.
Individual groups or members have received support from
RTP (Australia), 
FWO Odysseus grant G0ASD25N (Belgium), 
Key Research Program of Frontier Sciences of CAS, CAS PIFI, CAS CCEPP (China); 
Minciencias (Colombia);
EPLANET, Marie Sk\l{}odowska-Curie Actions, ERC and NextGenerationEU (European Union);
A*MIDEX, ANR, IPhU and Labex P2IO, and R\'{e}gion Auvergne-Rh\^{o}ne-Alpes (France);
Alexander-von-Humboldt Foundation (Germany);
ICSC (Italy); 
Severo Ochoa and Mar\'ia de Maeztu Units of Excellence, GVA, XuntaGal, GENCAT, InTalent-Inditex and Prog.~Atracci\'on Talento CM (Spain);
the Leverhulme Trust, the Royal Society and UKRI (United Kingdom).


\newpage

\clearpage

\section*{Appendices}
\appendix

\section{Correlation matrix}
\label{app:correlation}
\begin{figure}[h]
    \centering
    \includegraphics[width=0.9\linewidth]{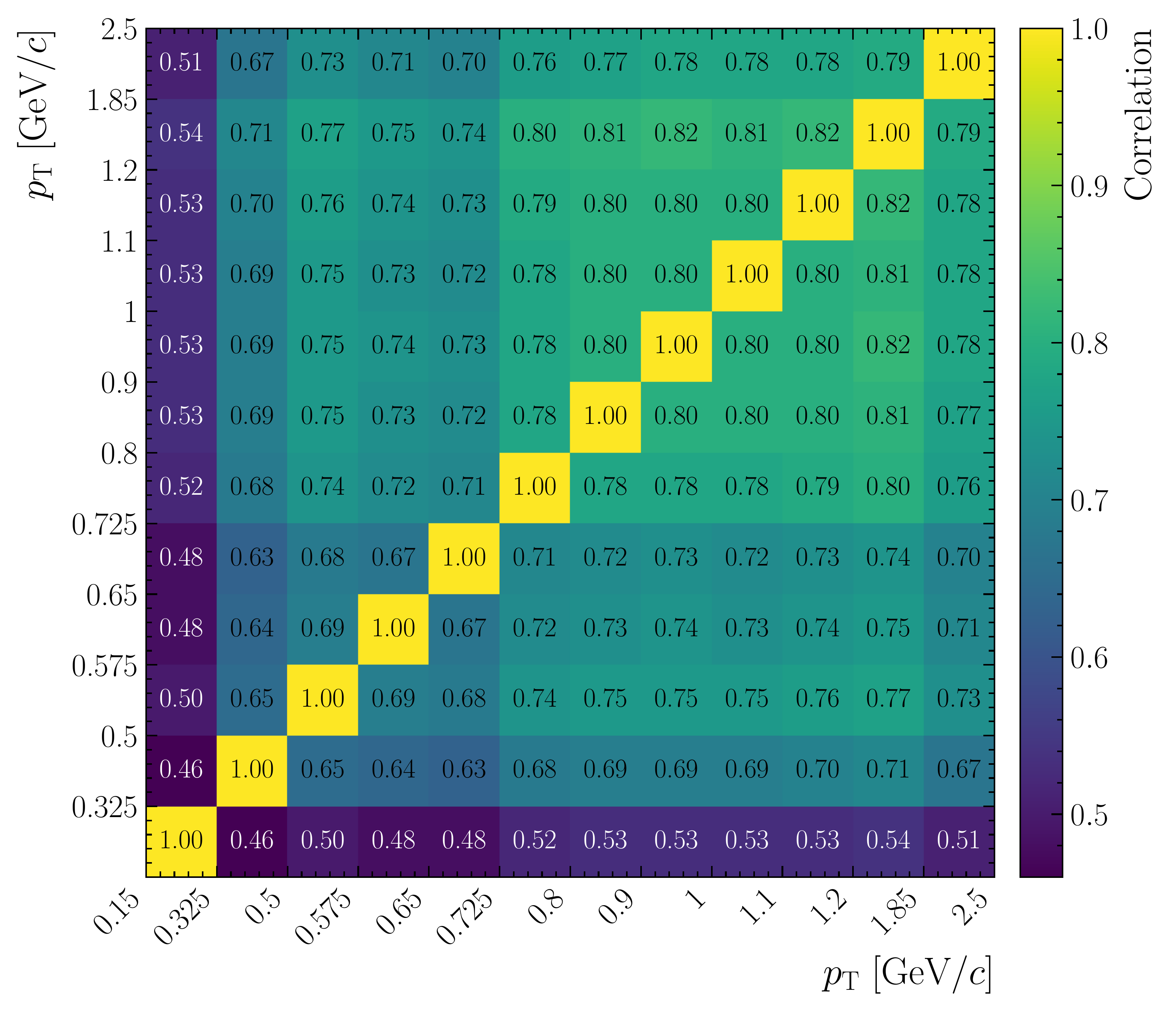}
    \caption{Correlation matrix of the results for the \pt binning. The matrix includes statistical and systematic uncertainties added in quadrature.}
    \label{fig:correlation_pT}
\end{figure}

\begin{figure}[h]
    \centering
    \includegraphics[width=0.9\linewidth]{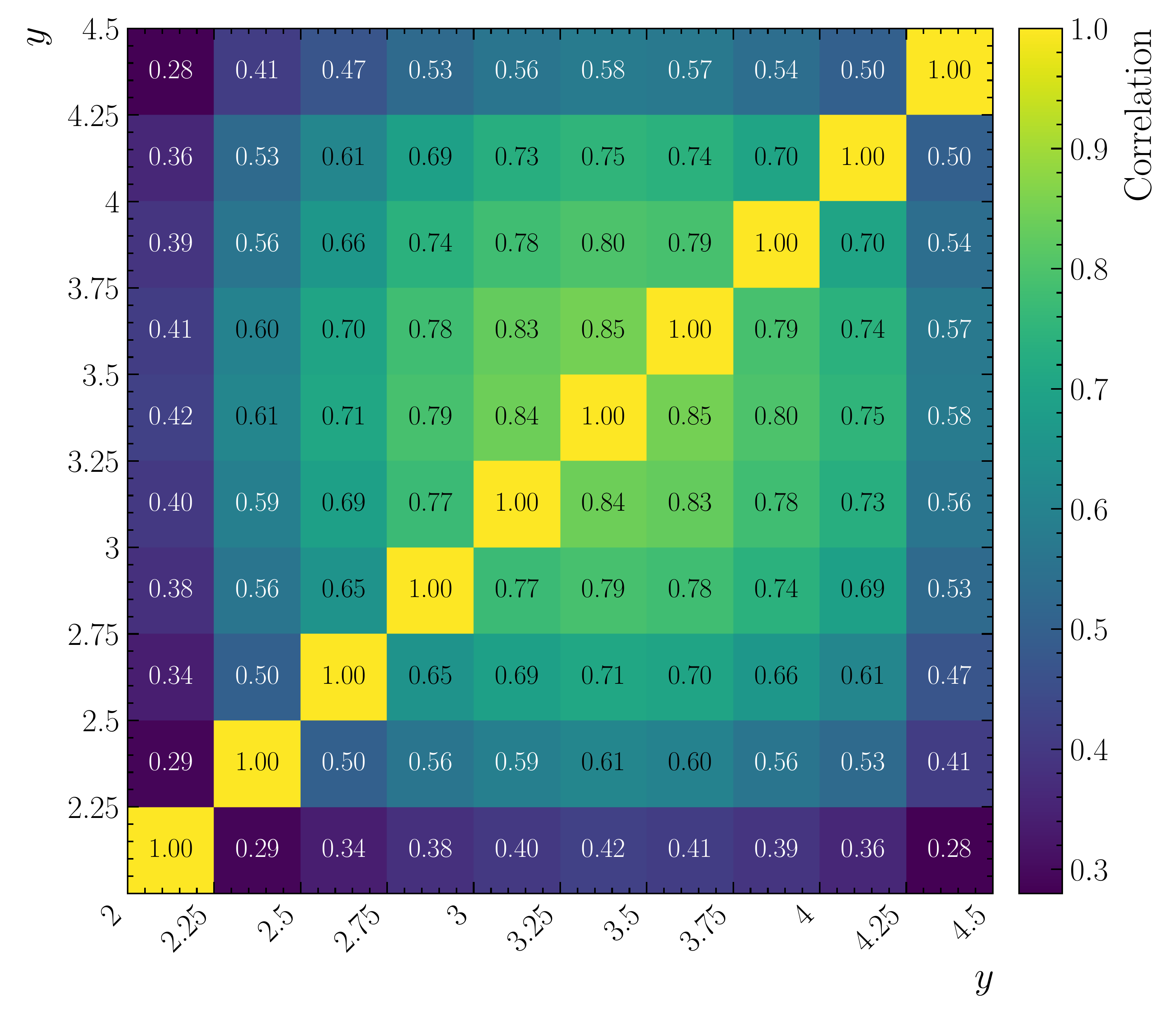}
    \caption{Correlation matrix of the results for the $y$ binning. The matrix includes statistical and systematic uncertainties added in quadrature.}
    \label{fig:correlation_y}
\end{figure}

Figures~\ref{fig:correlation_pT} and \ref{fig:correlation_y} provide the correlation matrices for the results in the in \pt and $y$ intervals.

\clearpage
\addcontentsline{toc}{section}{References}
\bibliographystyle{LHCb/LHCb}
\bibliography{main,LHCb/standard,LHCb/LHCb-PAPER,LHCb/LHCb-CONF,LHCb/LHCb-DP,LHCb/LHCb-TDR,LHCb/LHCb-PUB}

\newpage
\centerline
{\large\bf LHCb collaboration}
\begin
{flushleft}
\small
R.~Aaij$^{41}$\lhcborcid{0000-0003-0533-1952},
M.~Abdelfatah$^{74}$,
A.S.W.~Abdelmotteleb$^{62}$\lhcborcid{0000-0001-7905-0542},
C.~Abellan~Beteta$^{56}$\lhcborcid{0009-0009-0869-6798},
F.~Abudin\'en$^{64}$\lhcborcid{0000-0002-6737-3528},
T.~Ackernley$^{66}$\lhcborcid{0000-0002-5951-3498},
A.A.~Adefisoye$^{74}$\lhcborcid{0000-0003-2448-1550},
B.~Adeva$^{52}$\lhcborcid{0000-0001-9756-3712},
M.~Adinolfi$^{60}$\lhcborcid{0000-0002-1326-1264},
P.~Adlarson$^{91,46}$\lhcborcid{0000-0001-6280-3851}
% ...

{\footnotesize \it

% ...
$^{41}$Nikhef National Institute for Subatomic Physics, Amsterdam, Netherlands\\
% ...
$^{46}$National Center for Nuclear Research (NCBJ), Warsaw, Poland\\
% ...
$^{52}$Instituto Galego de F{\'\i}sica de Altas Enerx{\'\i}as (IGFAE), Universidade de Santiago de Compostela, Santiago de Compostela, Spain\\
% ...
$^{56}$Physik-Institut, Universit{\"a}t Z{\"u}rich, Z{\"u}rich, Switzerland\\
% ...
$^{60}$H.H. Wills Physics Laboratory, University of Bristol, Bristol, United Kingdom\\
% ...
$^{62}$Department of Physics, University of Warwick, Coventry, United Kingdom\\
% ...
$^{64}$School of Physics and Astronomy, University of Edinburgh, Edinburgh, United Kingdom\\
$^{65}$School of Physics and Astronomy, University of Glasgow, Glasgow, United Kingdom\\
$^{66}$Oliver Lodge Laboratory, University of Liverpool, Liverpool, United Kingdom\\
% ...
$^{74}$Syracuse University, Syracuse, NY, United States\\
% ...
$^{91}$Department of Physics and Astronomy, Uppsala University, Uppsala, Sweden, associated to $^{65}$\\
% ...
}
\end{flushleft}




\end{document}